# Deriving phosphorus atomic chain from few-layer black phosphorus

Zhangru Xiao[1#], Jingsi Qiao[2#], Wanglin Lu[1], Guojun Ye[3,4], Xianhui Chen[3,4], Ze Zhang[1], Wei Ji[2*], Jixue Li[1*] and Chuanhong Jin[1*]

[1]*State Key Laboratory of Silicon Materials, School of Materials Science and Engineering, Zhejiang University, Hangzhou, Zhejiang 310027, China*

[2]*Beijing Key Laboratory of Optoelectronic Functional Materials & Micro-Nano Devices, Department of Physics, Renmin University of China, Beijing 100872, China*

[3]*Key Laboratory of Strongly-coupled Quantum Matter Physics, Hefei National Laboratory for Physical Sciences at Microscale and Department of Physics, University of Science and Technology of China, Hefei, Anhui 230026, China,*

[4]*Collaborative Innovation Center of Advanced Microstructures, Nanjing University, Nanjing 210093, China*

[#] These authors contributed equally to this work.

**Email:** wji@ruc.edu.cn, jx_li@zju.edu.cn, chhjin@zju.edu.cn.

**Abstract**

Phosphorus atomic chains, the utmost-narrow nanostructures of black phosphorus (BP), are highly relevant to the in-depth development of BP into one-dimensional (1D) regime. In this contribution, we report a top-down route to prepare atomic chains of BP via electron beam sculpting inside a transmission electron microscope (TEM). The growth and dynamics (i.e. rupture and edge migration) of 1D phosphorus chains are experimentally captured for the first time. Furthermore, the dynamic behaviors and associated energetics of the as-formed phosphorus chains are further corroborated by density functional theory (DFT) calculations. The 1D counterpart of BP will serve as a novel platform and inspire further exploration of the versatile properties of BP.

**Introduction**

Black phosphorus [1-6], as the most stable allotrope of phosphorus [7], is a single-element crystalline material consisting of puckered atomic layers, coupled via weak inter-layer van der Waals interactions. Since its first demonstration as a channel material in a field effect transistor [1], BP has been regarded as an promising semiconductor for many potential applications in nanoelectronics and optoelectronics, largely owing to its excellent performance and fascinating properties [3, 8, 9]. Different from its single-elemental relative graphene in the family of two-dimensional (2D) materials, few- and mono-layer BP possesses unique highly anisotropic [10, 11] in-plane atomic and electronic structures, which further enriches the versatility of the properties of BP.

Quantum confinement will dramatically modulate the electronic structures of low-dimensional materials, and BP is not an exception. When its thickness is reduced from bulk to monolayer, the bandgap of BP increases from 0.3 eV to 2.0 eV [12, 13]. In respect to the confinement in the context of width, BP nanoribbons [14-17] are theoretically predicted to possess interesting properties including the orientation-dependent (either armchair or zigzag) scaling law and carrier mobility as a function of width. The introduction of external impurity (such as charge doping [18]) or edge defects [19, 20] could even tune the electronic structures and physical properties of these nanoribbons. Experimentally, BP nanoribbons are less studied partly due to the difficulty of sample preparation, until very recently Paul [21] et al. reported the successful fabrication of few-nanometer-width BP nanoribbons via electron sculpting.

As a BP nanoribbon further narrows down in width, it will enter the 1D regime, i.e. atomic chains of phosphorus, which are expected to exhibit specific electronic properties. Fabrication of the 1D atomic chain from BP will inspire the research on its atomic and electronic structures and promote the exploration of applications in future molecule devices. Experimentally, electron beam sculpting has been proved to be an effective way to achieve 1D atomic chains from 2D films counterparts in a number of systems, including graphene [22, 23], boron nitride (BN) [24] and transition-metal dichalcogenide (TMD) [25, 26]. From the theoretical side, Vierimaa [27] *et al* indicated that atomic chains of BP could be stable in certain configurations under electron beam illumination, while the BP atomic chains have not yet been realized experimentally.

To address this issue, we introduce here a top-down route to prepare phosphorus atomic chains via electron-beam sculpting. A combination of electron beam irradiation and thermal annealing was carried out to clean the as-exfoliated BP samples. Singlet and triplet atomic chains bridging the BP films were eventually fabricated. To reveal the atomic configuration of 1D chains, we also employ DFT calculations to explain our experimental findings.

**Methods**

Few-layer BP flakes were obtained by micro-mechanical exfoliation from bulk BP crystal. We adopted an all-dry method [28] to transfer the BP flakes onto holey $Si_3N_4$ membranes in ambient, and then loaded them into the TEM via a MEMS heating holder (DENSsolutions Inc.). As-exfoliated BP samples were kept in an argon gas-filling glovebox to minimize the ambient oxidation [29-31]. High-resolution TEM (HRTEM) imaging and structure engineering were conducted with a monochromated TEM (FEI Titan[80-300] equipped with an spherical aberration corrector in image side) operated at an acceleration voltage of 80 kV to decrease the radiation damage. HRTEM image simulations were done by Mactempas (Total resolution Inc.), in which the structures after total energy relaxation via DFT calculation are used as the input. Electron energy loss spectroscopy (EELS) was recorded by a GIF Quantum 965X (Gatan Inc.).

DFT calculations were performed using the generalized gradient approximation for the exchange-correlation potential, the projector augmented wave method [32, 33], and a plane wave basis set as implemented in the Vienna *ab-initio* simulation package (VASP) [34]. The energy cutoff for the plane-wave basis was set to 700 eV for infinite one dimensional atomic chain and 350 eV for supported atomic chain. A *k*-mesh of 1×4×1 was adopted to sample the first Brillouin zone of the 9×3 monolayer black phosphorus supercell (armchair direction (*x*) and zigzag direction (*y*)). In geometry optimization, van der Waals interactions were considered at the van der Waals density functional (vdW-DF) [35, 36] level with the optB86b functional for the exchange potential (optB86b-vdW) [37, 38]. Over 120 finite linear, armchair and zigzag chains models are fully relaxed considering different connection-edges (4-ring, 5-ring, 6-ring and so on), different directions (from [110] to [100]) and BP substrate effects. The shape and volume of each supercell were fixed and all atoms in it were allowed to relax until the residual force per atom was less than 0.01 eV·Å$^{-1}$. Energy per phosphorus (P) atom in finite chain supported by one layer (1L) BP is calculated with $E_p = (E_{tot} - E_{BP\text{-}ribbon})/N$, where $E_{tot}$ and $E_{BP\text{-}ribbon}$ are the total energy of the system and BP ribbons without chains respectively, and N is the number of P atoms in finite chains.

**Results and Discussions**

As-exfoliated BP sheets were firstly treated by electron beam irradiation ($10^5$ e/nm$^2$s) for about 10 minutes to decompose possible organic residuals introduced during the exfoliation and transfer process. Assisted by a thermal annealing treatment at about 523 K, a layer-by-layer sublimation process was realized to reduce the thickness of a BP sheet, as also demonstrated on graphite [22]. In this way, local regions with ultra-clean surface could be prepared *in-situ* on BP flakes, as clearly evidenced by the atomic-resolution TEM image shown in figure 1(a)-(c). The corresponding FFT (fast Fourier transform) map also confirms the crystallographic nature of the BP. For further studies, the electron dose was then reduced (~$10^4$ e/nm$^2$s) to minimize radiation damage.

The improvement in surface cleanliness is confirmed by EELS analysis before and after *in-situ* heating treatment, as shown in figure 1(d). The obvious oxygen K-edge recorded on as-exfoliated BP surface becomes negligible, after the combination of electron irradiation and *in-situ* heating process. The remaining presence of characteristic peaks including P $L_{2,3}$ and $L_1$ edges proves that such treatment did not change the BP nature of the sample.

Figure 2 presents time-consequential HRTEM images of the formation and breakage of the 1D phosphorus atomic chain during the layer-by-layer thinning process, see also movie S1 in the supporting information (SI). Firstly a BP nanoribbon (BPNR) with a width of 3.5 nm was fabricated with its edge aligned with the [100] direction (figure 2(a)). Since edge P atoms are predicted to have lower displacement threshold energy [27], the outermost P atoms of BPNR were preferentially displaced and after about 2.4 s the ribbon constricted to a 1D chain as shown in figure 2(b). The as-formed chain is about 2.5 nm in length and parallel to the [1-10] direction. It is noted that the chain is not suspended but supported by other BP layers. At about 4.8 s, the chain further develops into a longer one (about 5 nm) as a result of continuous sublimation of the BPNR as shown in figure 2(c). The left end of the chain (arrowed by a yellow triangle) migrates along the bridging edge on the BP flake, and realign itself from [1-10] to [110] direction. The intersection angle between the chain and the armchair direction remains the same (about 36°) before and after the migration, i.e. the two orientations are crystallographically equivalent. The chain survives for about 5 seconds before its breakage as shown in figure 2(d), where the black arrows point to the two ends of the chain shown in figure 2(c). Mechanical vibration, irradiation damage or heating-induced sublimation may be responsible for the breakage of the chain.

Figure 3 presents another case of the evolution of triplet phosphorus chains, marked as AA′, BB′ and CC′. Attached to the zigzag edge of the BP film, the chains are nearly parallel to each other and all are about 2.5 nm long, as shown in figure 3(a). After 4.8 s, the A, B and C ends of the triplet chains migrate simultaneously along the zigzag edge while the right-hand ends remain almost stationary (figure 3(b)). The migration behavior can be attributed to the release of the strain introduced by structure evolution. At 12 s and 14.4 s, the breakage of CC′ and BB′ was observed successively during the tilting process, and some black-spot like contrasts (identified by yellow arrows in figure 3(c) and 3(d)) appeared, which may correspond to a specific edge structure of BP or molecule clusters such as P4. Unfortunately, we were not able to resolve its atomic structure due to the influence of the supporting BP layers and the limited resolution of the TEM. At 16.8 s, AA′ is also broken and it is worth mentioning that the A end of AA′ chain (figure 3(d)) rebound to another place, identified by the white arrow in figure 3(e), which is unambiguously shown in movie S2 of SI. This behavior can be regarded as evidence for the release of the elastic energy stored by the atomic chain during its formation. The length of AA′ in figure 3(e) is almost the same as that in figure 3(d), so we can boldly predict that the breakpoint is near to the end of the chain. There are similar calculation results in carbon atomic chain [22], suggesting that the bonds of the atoms at the two ends are easier to break than others within the chain. The high activity of the connection between chain and edge also provides the possibility of the migration along the edge.

So far, we could not resolve the atomic structure of the as-formed phosphorus chain due to a few technical difficulties: (i) The e-beam irradiation damage on the chain and the BP layers is rather severe at 80 kV; (ii) The heating-induced sublimation is hard to control, so fabricating monolayer BP and suspended atomic chains intentionally is rather difficult; (iii) The BP substrate and the high flexibility of the chain make it difficult to identify its atomic structure by TEM.

DFT calculations were carried out to reveal likely atomic details for 1D phosphorus chains, and we considered infinite structure models first to minimize computing resources. Figure 4(a)-(e) shows the atomic structures of five representative infinite 1D phosphorus chains, including a linear chain, reconstructed zigzag (ZZ-R) chain, armchair (AC) chain, armchair and zigzag mixed (AC-ZZ) chain and zigzag (ZZ) chain. Table 1 summaries P-P bond angles and bond lengths of the five structure models together with the normalized total energy per P atom ($E_P$). It reveals that the stability order is ZZ, AC-ZZ, AC, ZZ-R and linear, similar to the previous calculation results [27]. Figure S1 (SI) shows three possible double-chain structure models, which are high-pressure phases and will not be discussed here. In terms of energetics, P-P bonds prefer roughly 100° folding angles rather than the 180° linear ones in 1D phosphorus chains, substantially different from the *sp* hybridization in carbon atomic chains, which is ascribed to the repulsion between two lone electron pairs of adjacent P atoms. For example, in a dimer of phosphinidene (phosphorus analog of carbene) the P atoms are arranged in zigzag pattern, as demonstrated by Liu *et al* [39].

To better match the experimental results, we also consider finite chains connected to BP edges and additionally supported by BP layers. The ends of finite chains are connected to zigzag edges in the all of the structure models. Local contact geometries of a finite chain and zigzag edge are discussed in SI (figure S2) and 6-ring (figure 5(a)) are the most energetically favored. The finite ZZ chain is predicted to be the most stable form, consistent with the conclusion from infinite chains. The ZZ chain is also fairly robust and can maintain its zigzag pattern with chain length varying from 6 Å to 30 Å and chain orientation rotated from [110] to [100] (figure S4, SI). Due to the stability of the chain in different directions, the experimentally observed migration of the chain in figure 3 case was indicated to be reasonable. Finite AC chains (figure S3(a), SI) are demonstrated to have strong interaction with supporting BP layers, and can only be stabilized along the [110] direction. But if supporting BP layers are removed, the finite AC chains prefer to keep their initial configurations (figure S5, SI). So the likely orientation preference along [110] direction observed in figure 2(b) and (c) should be related to the interaction between the chain and BP substrate. As for finite linear chains, they are highly unstable and tend to break into discrete P-P atoms.

Figure 5(a) shows a finite ZZ chain model with the length of roughly 2 nm, close to the length of experimentally observed phosphorus chains. In our structural relaxations, we used periodic boundary conditions and the model shown in figure 5(a) implies parallel chains with large-enough separations among them, which aims to simulate an isolated BP chain. Therefore, in our TEM image simulation, we only focus on the area marked by the black box in figure 5(b), with one complete chain and its neighboring chains removed, to help comparison between our experimental and simulated TEM images. Most of features of the phosphorus chain we observed in figure 2 are well reproduced by our simulated image shown in figure 5(b). To the best of our knowledge so far, the ZZ chain, as the most stable state theoretically predicted, is the best fitting to our experimentally recorded images.

**Conclusion**

To sum up, we have described a top-down route to prepare 1D phosphorus chains via e-beam sculpting at elevated temperature. For the first time，the dynamics of the formation, migration and breakage of the chain are captured in TEM images. DFT calculations and corresponding simulated image indicate that a ZZ chain matches best with the experimentally observed chain. Further exploration of the properties and fabrication of freestanding chains with well-controlled electron irradiation is anticipated in the future, and the chain can be expected to act as the thinnest connecting channel in BP based nano-electronics.

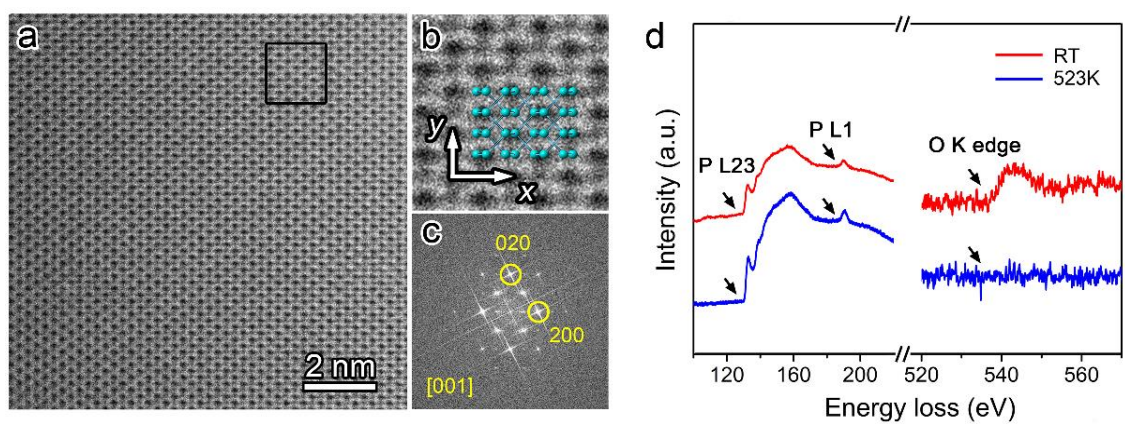

**Figure 1.** Structure and chemical analysis of BP sheets. (a) A representative HRTEM image of few-layer BP sheets with ultra-clean surface. (b) Amplified atomic-resolution HRTEM image of the marked area in (a). (c) The FFT image reveals the crystallographic information of BP. (d) EELS spectra shows the subtracted P $L_1$, $L_{23}$ and O K edge recorded on the BP samples before (red curve) and after (blue curve) beam irradiation and thermal annealing. No detectable oxygen peak was found on the EELS spectra recorded on the ultra-clean regions.

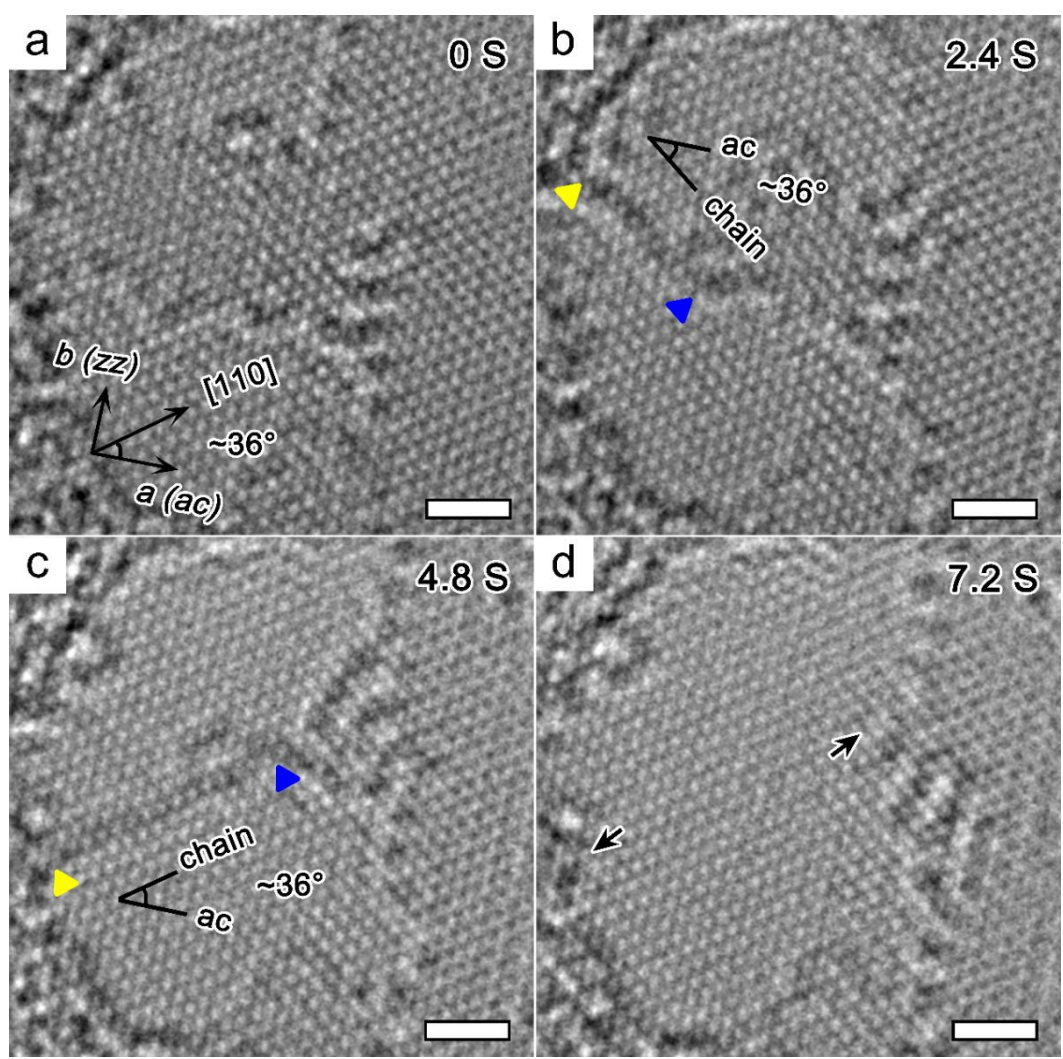

**Figure 2.** The formation of 1D phosphorus chain in BP. (a) A BP nanoribbon was fabricated under electron irradiation. (b) The phosphorus atoms were constantly removed, and a phosphorus chain of about 2.5 nm appeared. The ends of the chain were marked by a yellow triangle and a blue triangle

respectively. (c) One end of the chain pointed by the yellow arrow jumped to another place, and the chain became longer to 5 nm. The intersection angle between the chain and the armchair direction is both about 36° in (b) and (c). (d) The breakage of the chain. Scale bar is 2 nm in all panels.

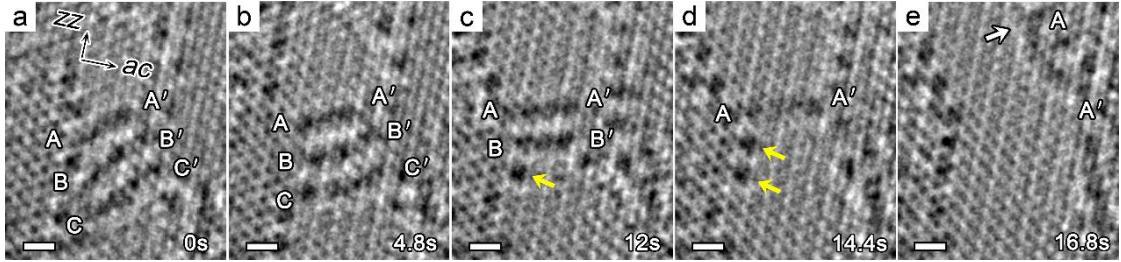

**Figure 3.** Dynamics of the 1D phosphorus chains in BP. (a) Triplet phosphorus chains appear with a large intersection angle with the armchair direction. (b) The left ends of the chains migrate along the zigzag edge, and the angle became smaller. (c) Double chains were left after the breakage of the CC′ chain. (d) A singlet chain was in a wide blurry contrast because of vibration. (e) The white arrow points at the A end of the singlet AA′ chain in (d), which is unambiguously shown in the movie S2 (SI). Scale bar is 1 nm in all panels.

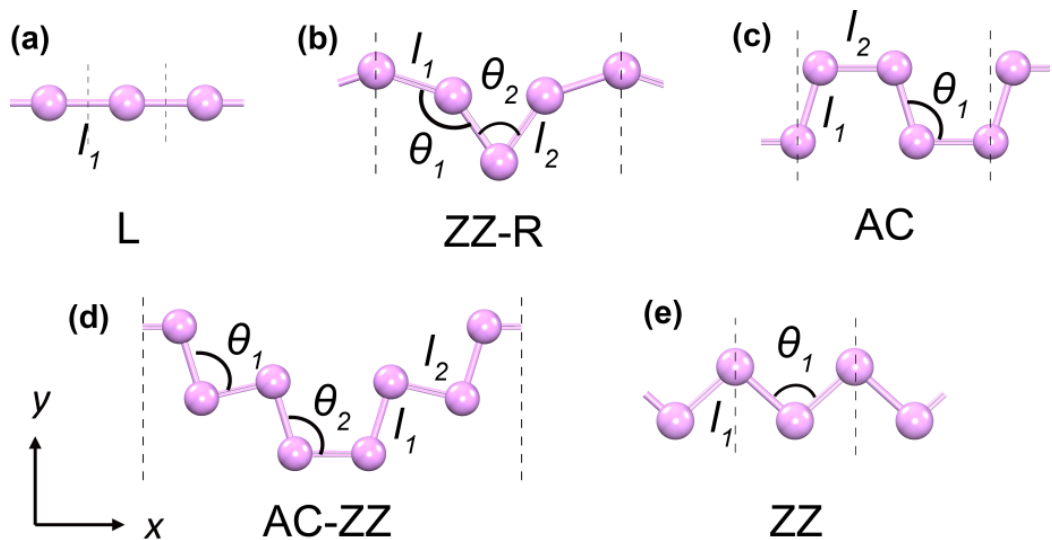

**Figure 4.** Top view of five kinds of representative infinite 1D phosphorus chains. (a) linear chain. (b) reconstructed zigzag (ZZ-R) chain. (c) armchair (AC) chain. (d) armchair and zigzag mixed

(AC-ZZ) chain. (e) zigzag (ZZ) chain. Normalized total energy per P atom ($E_P$), P-P bond angles and bond lengths of each configuration are shown in table 1.

**Table 1.** Single P atom energy ($E_P$), bond lengths ($l$), and bond angles ($\theta_1/\theta_2$) of atomic chains.

| System | $E_P$ (eV) | $l$ (Å) | $\theta_1/\theta_2$ (°) |
|---|---|---|---|
| Bulk BP | -4.11 | 2.24/2.27 | 96.1/101.6 |
| Linear | -2.67 | 2.13 | 180 |
| ZZ-R | -3.04 | 2.13/2.43 | 142.9/69.4 |
| AC | -3.37 | 2.09/2.18 | 106.7 |
| AC-ZZ | -3.42 | 2.16/2.10 | 93.7/107.0 |
| ZZ | -3.46 | 2.12 | 95.5 |
| Finite AC chain supported by 1L BP | -3.71 | 2.08/2.21 | 93.5~107.2 |
| Finite ZZ chain supported by 1L BP | -3.89 | 2.10/2.22 | 88.0~102.1 |

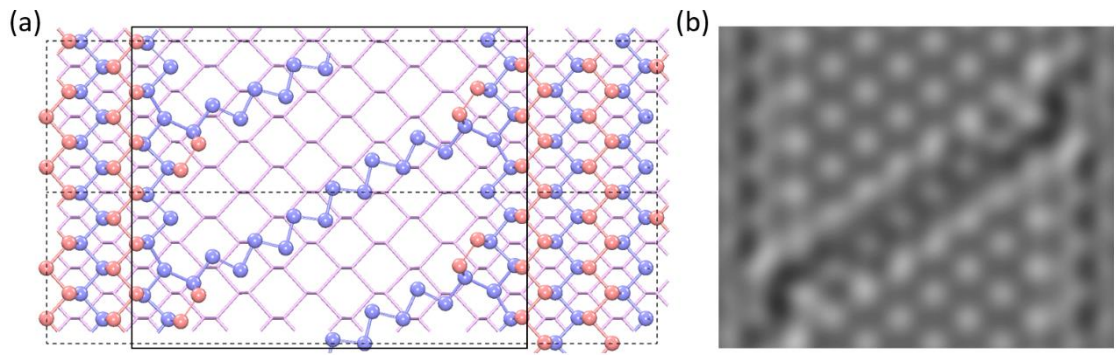

**Figure 5.** Geometric structure of finite ZZ chain. (a) Top view of finite 1D ZZ atomic chain supported by one layer BP. Side view would be found in SI (figure S3(b)). P atoms in different layer are clarified in different colors. (b) The simulated image of the model marked by the black box. Two halves of the chain without connection to each other are removed.

**Acknowledgement**

This work was financially supported by the National Basic Research Program of China (No. 2014CB932500 and No. 2015CB21004), the National Science Foundation of China (Nos. 51472215, 91433103, 11274380, 51222202, 11227403, 11534010) and the Fundamental Research Funds for the Central Universities under grant No. 16XNH062 (RUC). This work made use of the resources of the Center of Electron Microscopy of Zhejiang University. We thank Prof. Ray F. Egerton for fruitful discussions and suggestions.